\documentclass{pss}
\usepackage{epsf,pstricks}
\usepackage{amsmath}
\usepackage{epsfig,graphicx}
\begin{document}

\title{Horizontal line nodes in superconducting
Sr$_2$RuO$_4$}
\author{G. Litak$^{1,2}$, J.F. Annett$^3$, B.L. Gy\"{o}rffy$^3$, K.I.
Wysoki\'nski$^4$}
\address{
$^1$ Max-Planck-Institut f\"{u}r
Physik komplexer Systeme,
N\"{o}thnitzer Str. 38, \\
01187 Dresden, Germany
\\
$^2$Department of Mechanics, Technical University of Lublin,
Nadbystrzycka 36, \\
20-618 Lublin, Poland
\\
$^3$H.H. Wills Physics Laboratory, University of Bristol, Tyndall
Ave, Bristol, BS8 1TL, UK
\\
$^4$Institute of Physics, M. Curie -- Sk\l{}odowska University,
 ul. Radziszewskiego 10, \\ 20-031 Lublin, Poland}
\submitted{\today}
\maketitle

\hspace{9mm}Subject classification:74.70.Pq,74.20.Rp,74.25.Bt

\begin{abstract}
We analyze the possibilities of triplet pairing in Sr$_2$RuO$_4$
based upon an idea of interlayer coupling. 
We have considered two models differing by the 
 effective interactions. In one model the quasi-particle spectra  have
horizontal line nodes on all three Fermi surface sheets, while in
the other the spectra have line or point nodes on the $\alpha$ and
$\beta$ sheets and no nodes on the $\gamma$ sheet. Both models 
reproduce the experimental heat capacity and penetration depth results, 
but the 
calculated specific heat is sightly closer to experiment in the second
solution with nodes only on the $\alpha$ and $\beta$ sheets.
\end{abstract}

\section{1. Introduction.} ~\\

Strontium Ruthenate (Sr$_2$RuO$_4$) is widely believed to be a
spin triplet
superconductor\cite{Maeno_a,Maeno,Mackenzie03,Manske}, however the
theoretical model and particular pairing mechanism are still hotly
debated
\cite{hase00,kubo00,kuboki01,zhit,annett02,Wys03,koik03,annett03,hase03}.
Its lattice structure resembles the layered structure of the
cuprate La$_{2-x}$Ba$_x$CuO$_4$ but, instead of high temperature
d-wave superconductivity, the superconducting phase of strontium
ruthenate ($T_c \sim 1.5$K) appears to be  p-wave or f-wave in
nature. The strongest evidence for this comes from the $^{17}$O
NMR Knight shift data\cite{ishida98} and neutron scattering
experiments\cite{Duffy} which indicate that the in-plane Pauli
spin susceptibility is constant below $T_c$. These experiments
could be naturally explained with a triplet pairing state
 ${\bf d}(\bf k)={\bf
e}_z(k_x\pm ik_y)$
 in exact analogy with the ABM phase of superfluid $^3$He.
The $\mu$SR experiments\cite{luke} also indicate a spontaneous
breaking of time reversal symmetry at $T_c$, which would be
consistent with such a chiral ABM type state. On the other hand
several experiments\cite{NishiZaki,DalevH,tantar01} indicate that
the gap function must have lines of nodes on the Fermi surface,
unlike the simple ABM state which is nodeless on the three
cylindrical sheets, $\alpha$, $\beta$ and $\gamma$, of the Fermi
surface of Sr$_2$RuO$_4$.

In this paper we address the question of the possible location of
these line nodes on the  Fermi surface. We concentrate on the case
of horizontal lines of nodes, because vertical nodes (for example
in f-wave pairing states) appear to be inconsistent with the
absence of angular dependence of the thermal conductivity in an
a-b plane magnetic field\cite{tantar01,izawa01}. The presence of
horizontal lines of nodes, as originally suggested by Hasegawa,
Machida and Ohmi\cite{hase00}, cannot be explained in any 2-d
theoretical model but requires a 3-d model with at least some
component of the pairing  interaction acting between planes. A
number of such models assuming interlayer coupling have been
proposed\cite{kubo00,kuboki01,zhit,annett02,Wys03,koik03,annett03,hase03}.

The specific question which we address here is whether there are
horizontal line nodes on all three Fermi surface sheets, as
proposed recently by Koikegami, Yoshida and
Yanagisawa\cite{koik03}, or whether the nodes  are only on the
$\alpha$ and $\beta$ sheets, as proposed by Zhitomirsky and
Rice\cite{zhit} and in our earlier interlayer coupling
model\cite{annett02,Wys03,annett03}. To answer this question we
will analyze, an effective week coupling model where the
attractive interactions can appear between electrons on nearest
neighbour and next nearest neighbour lattice sites (symbolised by
2 and 3 lines in Fig. 1). We will contrast the predictions of our
original interlayer coupling model\cite{annett02,Wys03,annett03}
with ones chosen to reproduce the gap structure proposed by
Koikegami, Yoshida and Yanagisawa\cite{koik03}.

\begin{figure}[thb]
\centerline{\epsfig{file=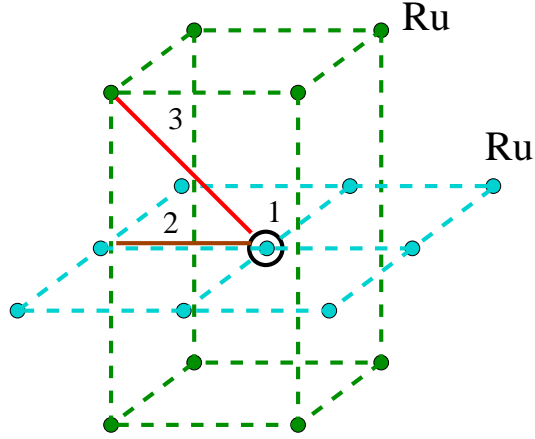,width=7cm,angle=0}}
\caption{Body-centred tetragonal lattice of Ruthenium atoms in the
Sr$_2$RuO$_4$ structure. The full lines show possible interactions
between electrons occupying the single site '1' and its in-plane
nearest neighbour sites '2' and intra-plane neighbours '3'.}
\label{fig1}
\end{figure}

\begin{figure}[h]
\epsfxsize=7cm
\centerline{\epsfig{file=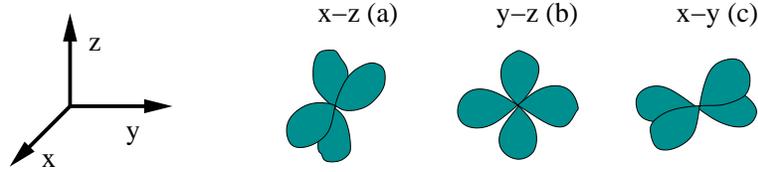,width=10.0cm,angle=0}}
\caption{Orientation of
$d_{xz}$ and $d_{yz}$ and $d_{xy}$
orbitals.}
\end{figure}

\begin{figure}[h]
\epsfig{file=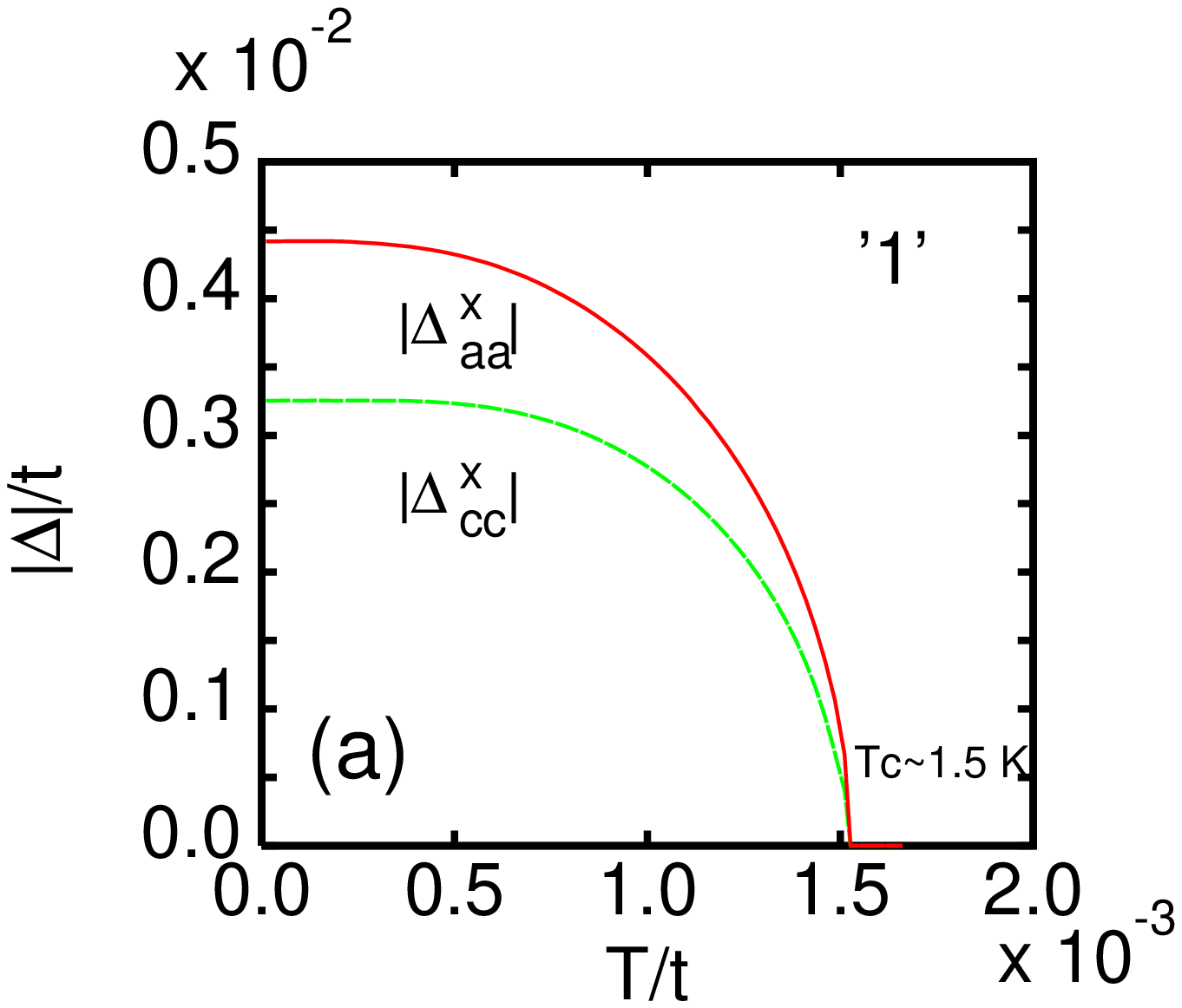,width=5.5cm,angle=-90} \hspace{-1.0cm}
\epsfig{file=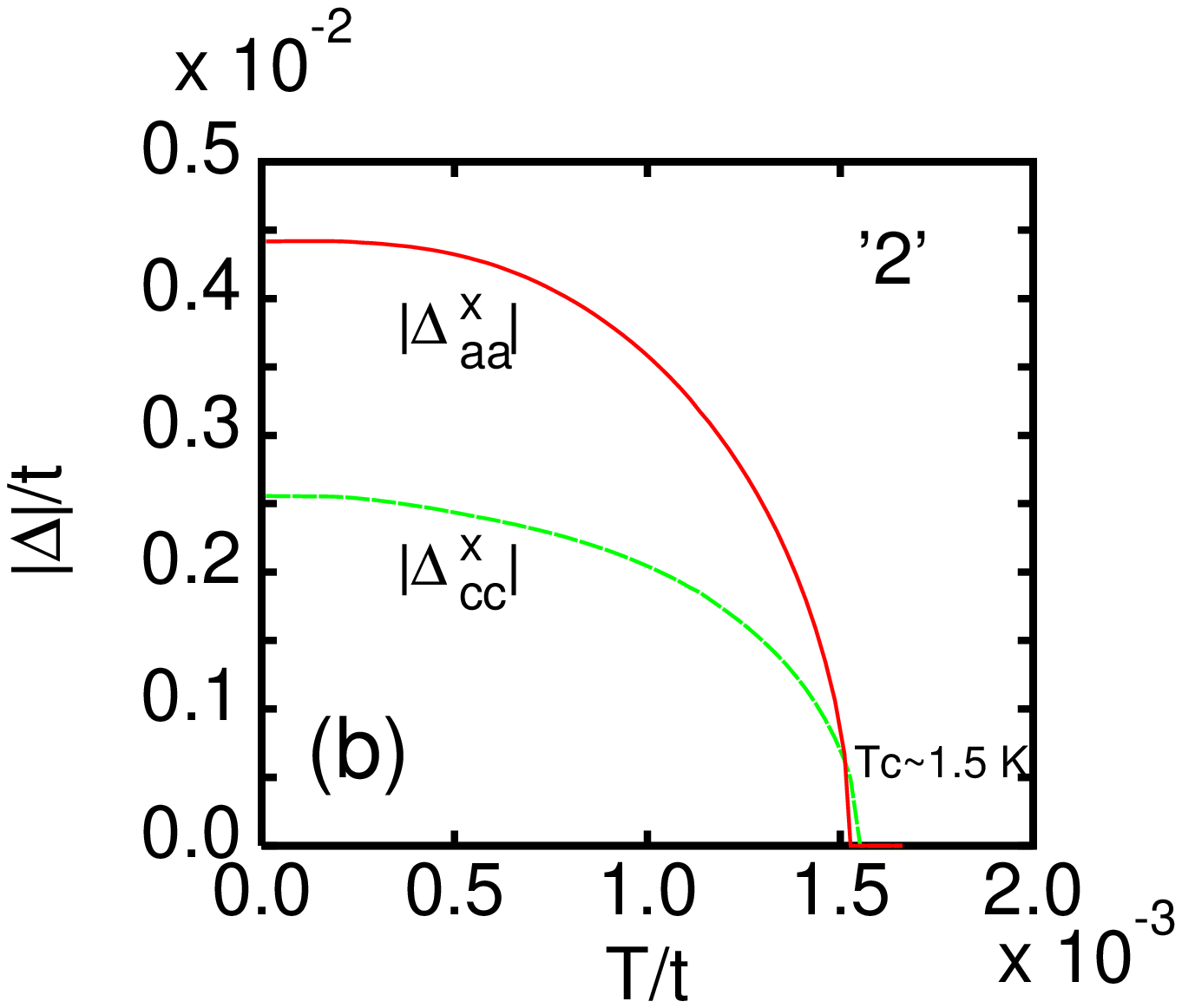,width=5.5cm,angle=-90}
\vspace{0.5cm}
\caption{
Temperature dependence of order parameters
$\Delta_{aa}^{\perp,x}$,  $\Delta_{cc}^{\perp,x}$ (a)
and $\Delta_{aa}^{\perp,x}$ and $\Delta_{cc}^{\parallel,x}$ (b)
found for different choice of interactions.
Fig. 3a corresponds to
the results for only
$U^{\perp}_{m,m'}(ij) < 0$ (Eq. 2 - scenario 1)
while
Fig. 3b to the case where  $U^{\perp}_{m,m'}(ij) < 0$ for
$m,m'=a,b$ (Eq. 3 - scenario 2) and
$U^{\parallel}_{c,c}(ij) < 0$.
}
\label{fig2}
\end{figure}

\begin{figure}[h]
\epsfig{file=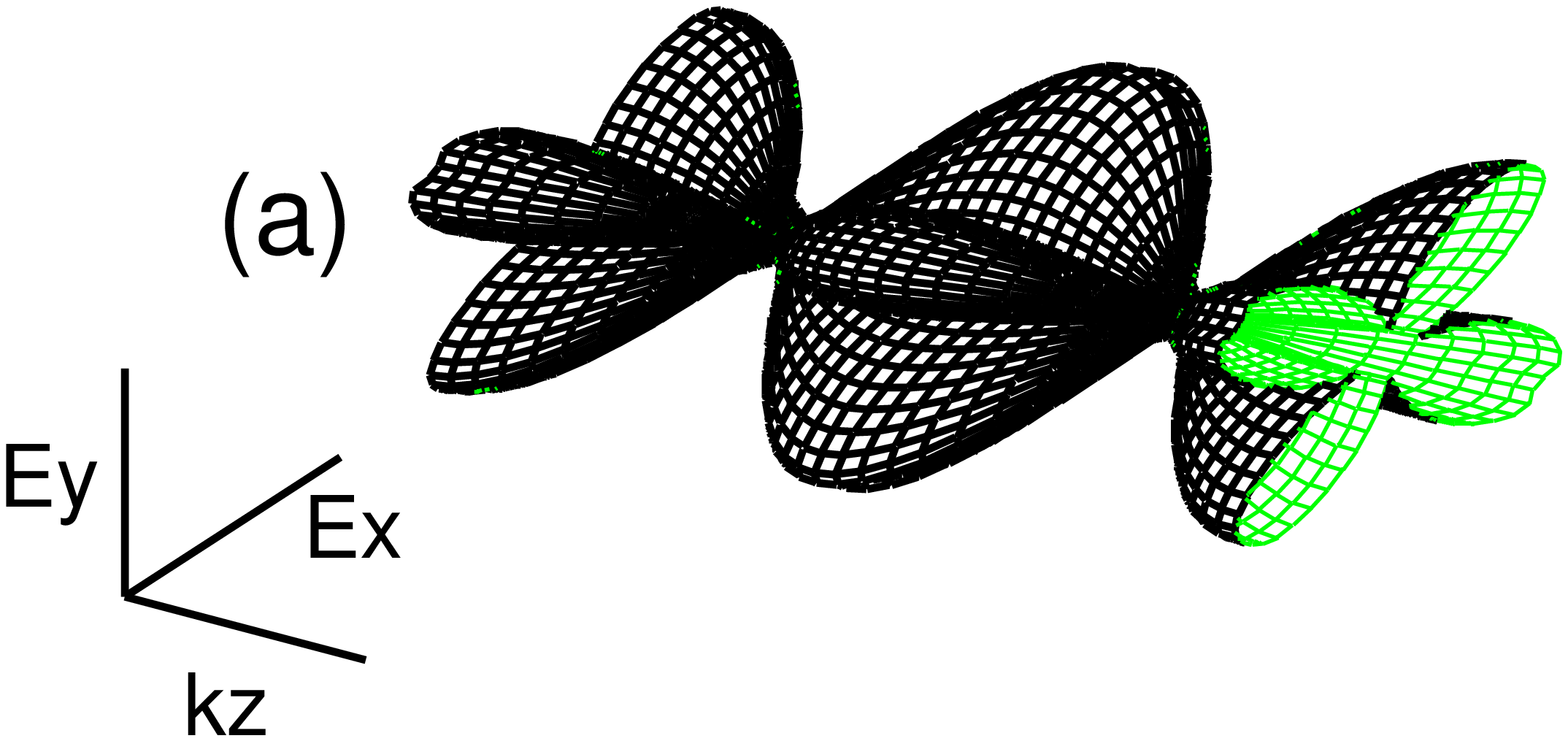,width=5.0cm,angle=-90} \hspace{-1cm}
\epsfig{file=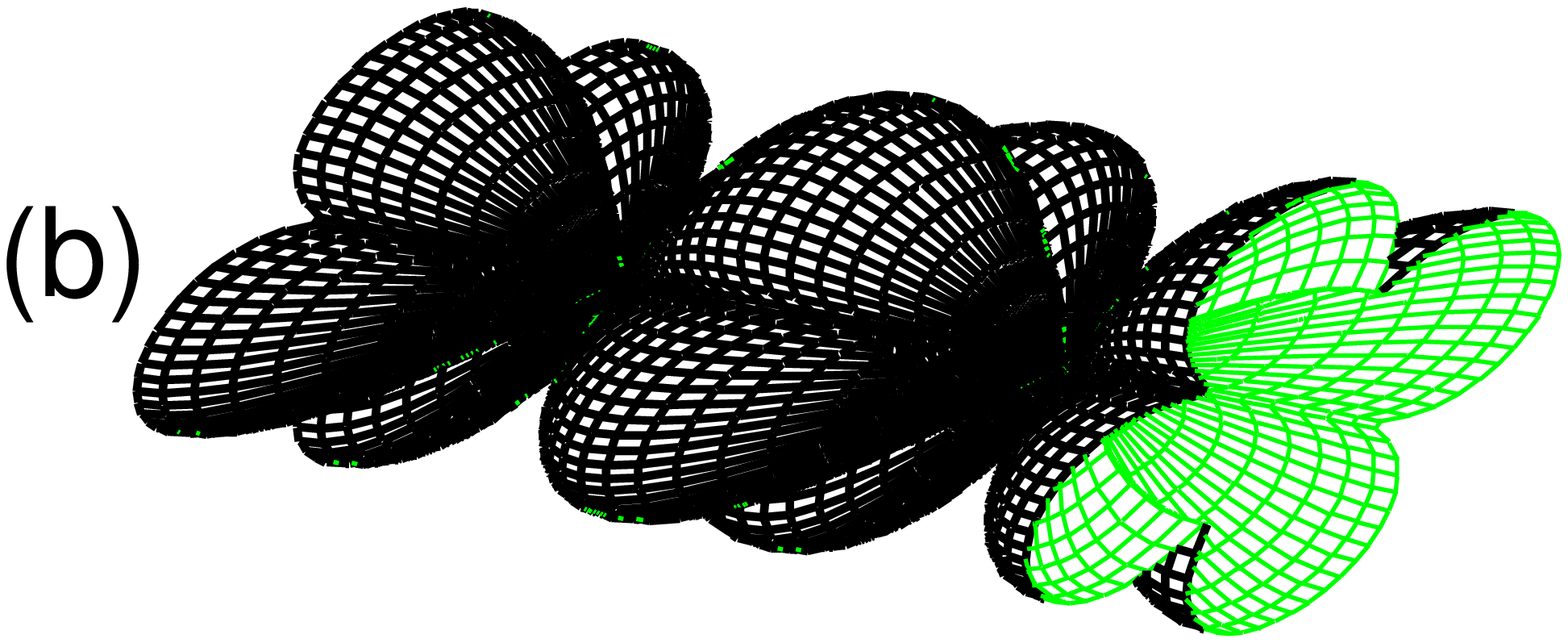,width=5.0cm,angle=-90}
\vspace{-3cm}

\epsfig{file=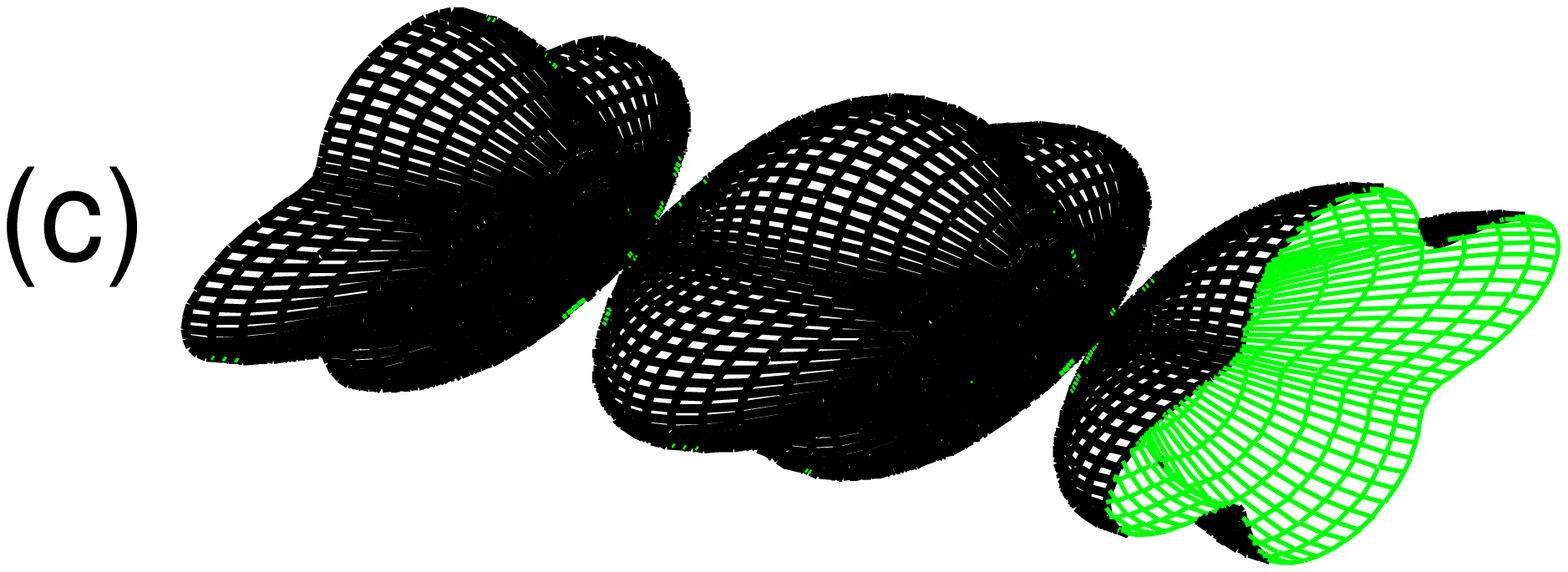,width=5.0cm,angle=-90} \hspace{-1cm}
\epsfig{file=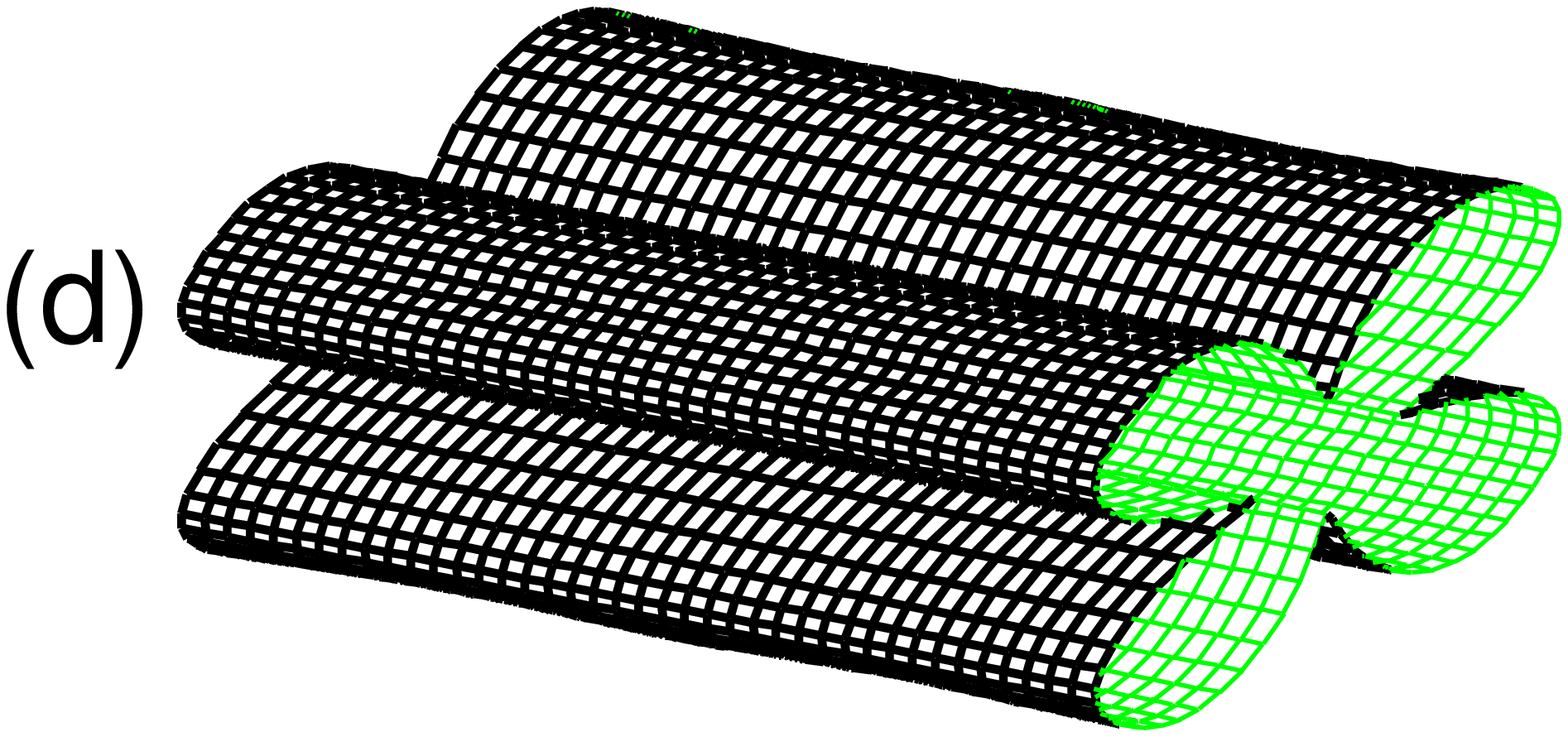,width=5.0cm,angle=-90}

\caption{Minimum energy quasiparticle eigenvalues on the $\alpha$ (a) ,
$\beta$ (b) and $\gamma$ (c,d) Fermi surface sheets. Fig. 4c 
(scenario 1) corresponds to
the results for only
$U^{\perp}_{m,m'}(ij) < 0$
while
Fig. 4d (scenario 2) to the case where  $U^{\perp}_{m,m'}(ij) < 0$ for
$m,m'=a,b$ and
$U^{\parallel}_{c,c}(ij) < 0$. Fig. 4a and b are the same for both cases.
}
\end{figure}

\begin{figure}[h]
\centerline{\epsfig{file=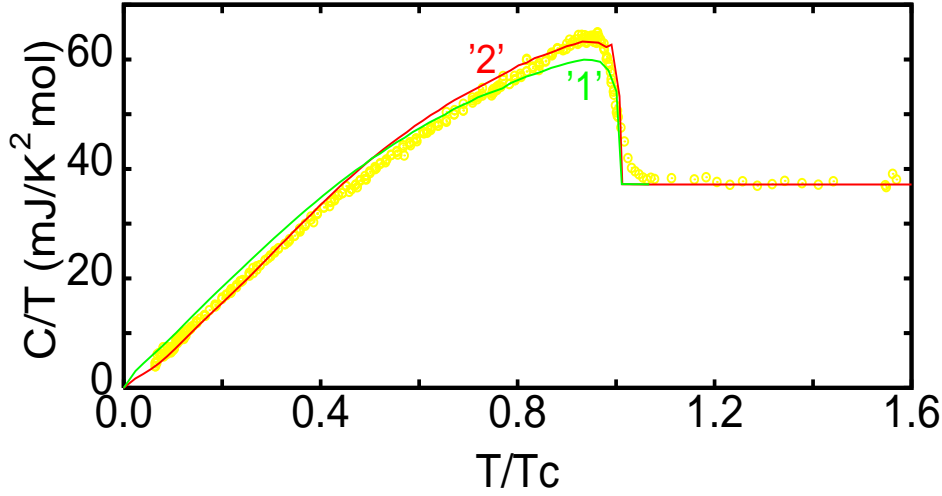,width=7.0cm,angle=-90}}
\vspace{0cm} \caption{Calculated specific heat, $C$, as a function
of temperature, $T$, compared to the experimental data of
NishiZaki {\it et al.}\cite{NishiZaki} (gray circles). The line '1'
shows the results for only
$U^{\perp}_{m,m'}(ij) < 0$
while
line '2' corresponds to the case where  $U^{\perp}_{m,m'}(ij) < 0$ for
$m,m'=a,b$ and
$U^{\parallel}_{c,c}(ij) < 0$.
}
\label{specific}
\end{figure}


\section{2. The interlayer coupling model} ~\\

\begin{figure}[h]
\centerline{\epsfig{file=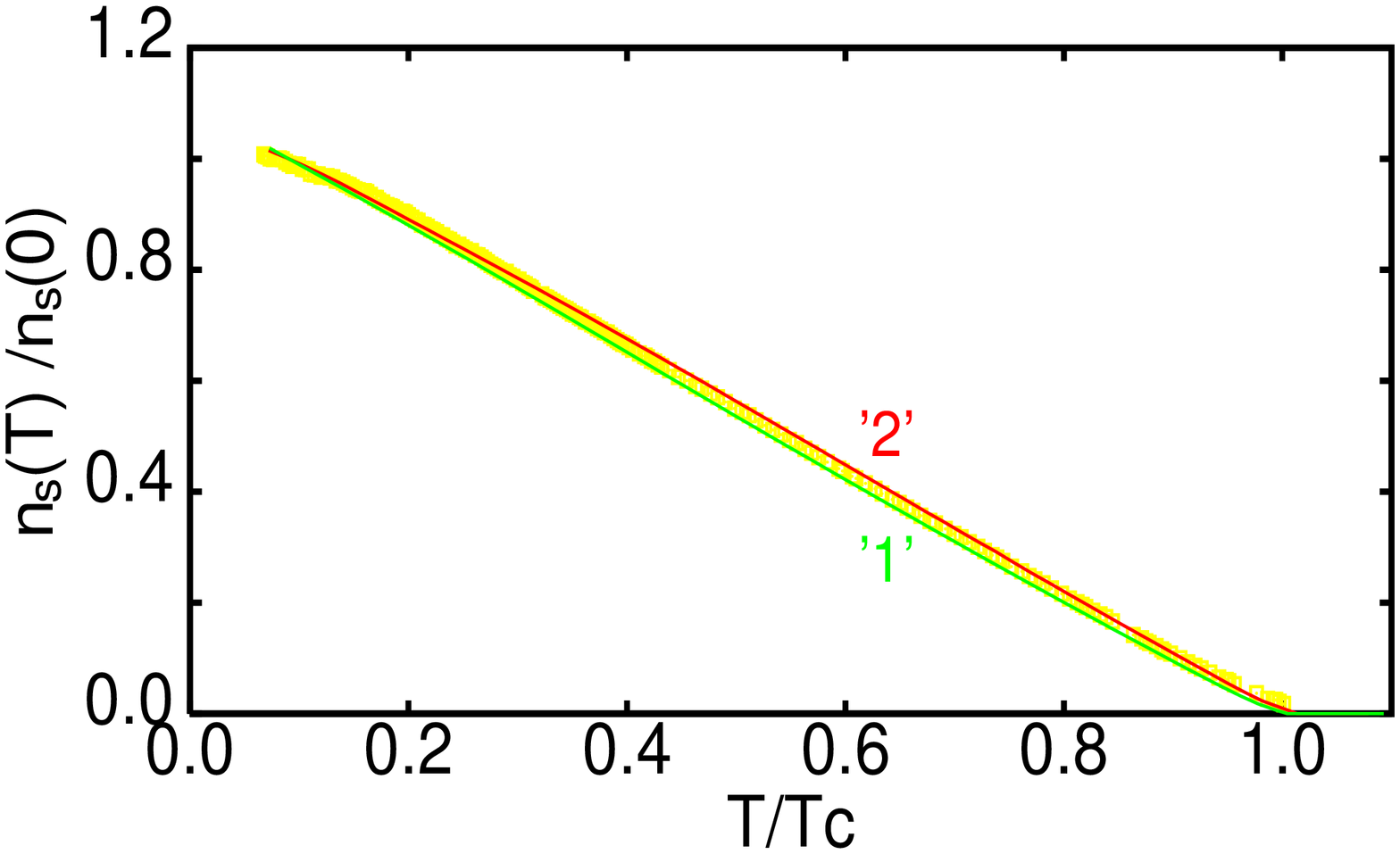,width=7.0cm,angle=-90}}
\vspace{0cm} \caption{Calculated superfuid density, $n_s$, as a function
of temperature, $T$, compared to the experimental data of
Bonalde {\it et al.}\cite{DalevH} (gray squares). The line '1'
shows the results for only
$U^{\perp}_{m,m'}(ij) < 0$
while
line '2' corresponds to the case where  $U^{\perp}_{m,m'}(ij) < 0$ for
$m,m'=a,b$ and
$U^{\parallel}_{c,c}(ij) < 0$.
}
\label{specific}
\end{figure}

 To describe superconductivity in
Sr$_2$RuO$_4$, we start from the following simple multi-orbital
attractive Hubbard Hamiltonian,
\begin{eqnarray}
  \hat{H}& =& \sum_{ijmm',\sigma}
\left( (\varepsilon_m  - \mu)\delta_{ij}\delta_{mm'}
 - t_{mm'}(ij) \right) \hat{c}^+_{im\sigma}\hat{c}_{jm'\sigma} \nonumber \\
&& - \frac{1}{2} \sum_{ijmm'\sigma\sigma'} U_{mm'}^{\sigma\sigma'}(ij)
 \hat{n}_{im\sigma}\hat{n}_{jm'\sigma'} .\label{hubbard}
\end{eqnarray}
Here $i$ and $j$ label the sites of a body centred tetragonal
lattice (as shown in Fig. 1),  and $m$ and $m^{\prime }$ refer to
the three Ruthenium $t_{2g}$ orbitals (as shown in Fig. 2). In the
following the orbitals will be denoted $a=xz$, $b=yz$ and $c=xy$.
The hopping integrals $t_{mm^{\prime }}(ij)$ and site energies
$\varepsilon _{m}$ were fitted to reproduce the experimentally
determined three-dimensional Fermi surface~\cite{mac,berg}. We
found that the set {$t_{mm'}$} given in \cite{annett03}gave a good
account of the Fermi surface data.


Since the actual physical mechanism of pairing in Sr$_2$RuO$_4$ is
unknown, we will adopt a phenomenological approach and treat the
effective Hubbard interaction constants
$U_{mm'}^{\sigma\sigma'}(ij)$ as free parameters. The idea is to
test different ``scenarios'' for the interaction constants against
the available experimental results.  Good agreement between
experiment and theory is likely to only occur when the interaction
parameters $U_{mm'}^{\sigma\sigma'}(ij)$ lead to a gap function on
the Fermi surface ${\bf d}({\bf k})$ which is similar to that
actually present in the material. In particular, both the jump in
specific heat at $T_c$, and the linear dependence of $C(T)/T$
($C(T)/T \sim T$ for line nodes) near to $T=0$ are sensitive to
the gap ${\bf d}({\bf k})$  over the whole Fermi surface.
Scenarios in which the line nodes occur on all Fermi surface
sheets or only on $\alpha$ $\beta$ would be expected to lead to
different temperature dependencies of $C(T)$, which can therefore
 be distinguished by comparison to the experiments.

In Fig. 1 we depict schematically the possible interactions
$U_{mm'}^{\sigma\sigma'}(ij)$ between electrons on the same
Ru-atom (shown by circle '1') and the neighbouring  Ru-atoms
(shown by lines denoted '2','3' for in-plane and out of plane
interactions, respectively). We assume that the dominant
interaction between electrons on the same site $i$ ('1') is the
strong Coulomb repulsion.  The effective interactions between
nearest neighbours are assumed to be attractive. They could arise
either from spin-fluctuation mediated exchange\cite{Manske}, or
from interlayer Coulomb scattering\cite{koik03}.

In this paper we will  compare two different model sets of
interaction constants.  The first  is motivated by the inter-plane
Coulomb scattering model of Koikegami, Yoshida and
Yanagisawa\cite{koik03}. It assumes Coulomb repulsion between
electrons on the same and the nearest neighbour in plane lattice
sites ('1' and '2' in Fig. 1) and attraction between Ru planes
('3' in Fig. 1).  The corresponding Hubbard  interaction
parameters are
\begin{equation}
 {\bf U}^\parallel =
\left(\begin{array}{ccc}
      0 & 0 & 0 \\
      0 & 0 & 0 \\
      0 & 0 & 0
      \end{array}
      \right)~~~~~~{\rm and}~~~~~~
    {\bf U}^\perp  =
          \left(
          \begin{array}{ccc}
           U_{\perp} & U^{\perp} & 0 \\
           U_{\perp} & U^{\perp} & 0 \\
           0 & 0 & U'^{\perp}
         \end{array}\right),
\end{equation}
expressed as matrices in orbital space, $m$, $m'$. In this model
there are two attractive Hubbard parameters, one acting only
between $c$ orbitals  and the other acting equally between $a$ and
$b$ orbitals. In k-space there is little hybridization between the
$a$-$b$ orbitals and the $c$ orbitals, and so these inter-plane
interactions mainly corresponding to interactions within the
$\gamma$ band ($U'^{\perp}$) and within the  $\alpha$ and $\beta$
Fermi surface sheets  ($U_{\perp}$). We shall call this case 'scenario 1'
in the rest of this paper.

We wish to compare the predictions of the above set of model
parameters with those of our previous inter-plane coupling
model\cite{annett02,Wys03,annett03}, which was motivated by the
different spatial orientations of the xz, yz, xy orbitals as shown
in Fig. 2. Given that the Ru d-xy orbital ($c$) has a mainly 2-d
character we assumed that interactions between $c$ orbitals are
mainly in-plane.  On the other hand the Ru d-xz (a) and d-yz (b)
orbitals are oriented perpendicular to the RuO$_2$ plane, and so
we assumed that the interactions between electrons in these
orbitals are mainly iner-plane.  This simple reasoning leads to
the following two parameter Hubbard model with
\begin{equation}
{\bf U}^\parallel =
\left(\begin{array}{ccc}
      0 & 0 & 0 \\
      0 & 0 & 0 \\
      0 & 0 & U^{\parallel}
      \end{array}
      \right)~~~~~~{\rm and}~~~~~~
   {\bf U}^\perp  =
          \left(
          \begin{array}{ccc}
           U^{\perp} & U^{\perp} & 0 \\
           U^{\perp} & U^{\perp} & 0 \\
           0 & 0 & 0
         \end{array}\right).
\end{equation}
We shall refer to this as 'scenario 2' below.

In our earlier papers\cite{annett02,Wys03,annett03} we showed that
this simple model gives a good overall account of the temperature
dependent heat capacity $C(T)$, in-plane superfluid density
$n_s(T)$, and thermal conductivity.  In \cite{annett03} we showed
that the predictions of the model are fairly robust against the
addition of extra interaction parameters or disorder.

In equations 2 and 3 we have set to zero any interaction terms
which are either assumed to be small, or those which may be
repulsive.  We have checked that all the zero values appearing in
the above interaction matrices ${\bf U}^{\parallel}$ and ${\bf
U}^{\perp}$ (Eqs. 2,3) can be changed into small positive values
representing repulsions without any change to the solution, and so we
decided to use (Eqs. 2,3) the minimal set leading to pairing.


For above choices of interactions within the negative $U$ extended
Hubbard model (Eq. 1), we
solved  the
Bogolubov-de Gennes
equations:
\begin{equation}
 \sum_{jm'\sigma'}
 \left(
  \begin{array}{ll}
  E^\nu - H_{m,m'}(ij) & \Delta^{\sigma\sigma'}_{m,m'}(ij) \\
  \Delta^{*\sigma\sigma'}_{mm'}(ij) & E^\nu + H_{mm'}
  \end{array}
  \right)
  \left(
  \begin{array}{l}
   u^\nu_{jm\sigma'}\\
   v^\nu_{jm'\sigma'}
 \end{array}\right) = 0 \label{eq:bdegequation}
\end{equation}
together with the self-consistency condition
\begin{equation}
\Delta^{\sigma\sigma'}_{mm'} = U^{\sigma\sigma'}_{mm'}(ij)
 \chi^{\sigma\sigma}_{mm'} (ij);~~~~~
\chi^{\sigma\sigma'}_{mm'}(ij) = \sum_{\nu} u^\nu_{im\sigma}
v^{*\nu}_{jm'\sigma'}(1 - 2f(E^\nu)),
\end{equation}
which follow from Eq. (1) on making the  usual BCS-like mean field
approximation\cite{ketterson}. Here $f(E^\nu)$
 is  the Fermi function,
$\beta=1/k_BT$, $k_B$ is Boltzmann constant and $\nu$ enumerates
the solutions of Eq.~\ref{eq:bdegequation}.

Assuming an   p-wave pairing state of the form ${\bf d}({\bf k})
\sim {\bf e}_z$, on each Fermi surface sheet, then we need only
consider the gap parameters $\Delta^{\uparrow\downarrow}_{mm'}(\bf
k)$ at each point in the Brillouin zone. Dropping the spin indices
for clarity, the general structure of pairing parameter  is of the
general form
\begin{eqnarray}
&& \Delta_{mm'}({\bf k}) = \Delta^x_{mm'} \sin{k_x} +
\Delta^y_{mm'} \sin{k_y } \nonumber\\
&& + \Delta^z_{mm'}\sin{k_z c\over 2} \cos{k_x\over 2} \cos{k_y
\over 2} + \Delta^f_{mm'} \sin{k_x \over 2} \sin{k_y \over 2}
\sin{k_z c\over 2} \nonumber
\\
&& + \left( \Delta^x_{mm'} \sin {k_x\over 2} \cos {k_y\over 2}
+ \Delta^y_{mm'} \cos{k_y \over 2} \sin {k_y\over 2}\right)
  \cos{k_z c\over 2}
\end{eqnarray}
for $m,m' = a$, $b$ and $c$. In the present calculations we
neglected the possibilities of $p_z$ pairing ($\Delta^z_{mm'}$) or
f-wave pairing $\Delta^f_{mm'}$, for reasons which are discussed
further in \cite{annett02,annett03}.

\section{3. Line Nodes and Specific Heat} ~\\

From the two different interaction models given by Eqs. 2 and 3 we
numerically find the corresponding solutions of the gap equation
Eqs. 4, 5. In the case of scenario '1', Eq. 2, the gap
parameters have the general form,
\begin{eqnarray}
 \Delta_{mm'}({\bf k}) = \left( \Delta^x_{mm'} \sin {k_x\over 2} \cos {k_y\over 2}
+ \Delta^y_{mm'} \cos{k_y \over 2} \sin {k_y\over 2}\right)
  \cos{k_z c\over 2},
\end{eqnarray}
for $m,m'=a,b,c$.  While for scenario '2'  (Eq. 3) the gap
parameters are
\begin{eqnarray}
&& \Delta_{cc}({\bf k}) = \Delta^x_{cc} \sin k_x +
\Delta^y_{cc} \sin k_y  \nonumber\\
&& \Delta_{mm'}({\bf k}) =
 \left( \Delta^x_{mm'} \sin {k_x\over 2} \cos {k_y\over 2}
+ \Delta^y_{mm'} \cos{k_y \over 2} \sin {k_y\over 2}\right)
  \cos{k_z c\over 2}
\end{eqnarray}
for $m,m' = a$ or $b$, respectively. Clearly scenario '1' has a
order parameter for which all of the gap parameters
$\Delta_{mm'}({\bf k})$ vanish in the planes $k_z = \pm \pi/c 2$.
Therefore all three Fermi surface sheets should have horizontal
line nodes. On the other hand, in scenario '2' only the $a$ and
$b$ components of $\Delta_{mm'}({\bf k})$ vanish, implying that
the $\gamma$ sheet is nodeless.

The temperature dependence of the order parameters in each
scenario are shown in Fig. 3. In our analysis we have fitted the
interaction parameters ($U^{\parallel}=-0.494t$,
$U^{\perp}=-0.590t$, $U'^{\perp}=-0.312t$)  to obtain a single
critical temperature $T_c \approx 1.5$K for all order parameter
components (Fig. 3). Note, the differences between
$U^{\perp}=-0.590t$ and $U'^{\perp}=-0.312t$ arises as an effect
of the difference in partial density of states for the bands
$\alpha$, $\beta$ and $\gamma$. In case of $\gamma$ the Fermi
surface is close to a Van Hove singularity \cite{koik03,litak00},
and so  $U'^{\perp}$ can be smaller while still obtaining the same
$T_c$. On the other hand, the larger  value of $U^{\perp}$ can be
explained by the out of plane spatial orientation of the $a$ and
$b$ Ru orbitals, compared to  the in plane Ru d-xy orientation of
$c$ orbital

The angular dependence of the eigenvalues  $E_{k_i, \rho}$, $i=x$,
$y$ on the corresponding $\rho= \alpha$, $\beta$, $\gamma$ Fermi
surface sheets, plotted along the $z$ axis, are shown in Fig. 4.
In case of the inter-plane attraction only scenario '1' the gap
has line nodes on all three  Fermi surface sheets. In contrast, in
scenario '2'  the gap is nodeless on the $\gamma$ sheet, as can be
seen in Fig 4(d).

Now the key question is whether experiment can distinguish between
these two gap scenarios, namely line nodes on all Fermi surface
sheets compared to just nodes on  $\alpha$ $\beta$ only. We
calculated the specific heat  for those solutions via the
following relation:
\begin{equation}
C = -2  k_B \beta^2 \frac{1}{N} \sum_{{\bf \rm k}, \rho} E_{{\bf \rm k} 
\rho} \frac{
\partial f (E_{{\bf \rm k} \rho}) }{\partial \beta},
\end{equation}
where $f$ is the Fermi function.

The results are  presented in Fig. 5 and compared to the
experimental values of Nishizaki {\it et al.}\cite{NishiZaki}.

One can see in Fig. 5 that the slope of $C(T)/T$ near to $T=0$ is
slightly higher for scenario '1' compared to scenario '2', consistent
with the ''extra'' line node on the $\gamma$ Fermi surface sheet.
However the change in slope is quite small, and so one can say
that either model is consistent with the low temperature
experimental data.

On the other hand, in Fig. 5 it is clear that the second solution
'2'  works slightly better for the jump of specific heat at
critical temperature $T_c$. Solution '1' has a smaller Fermi surface
average of $|\Delta({\bf k})|^2$, and this leads to the slightly
smaller jump in specific heat at $T_c$. The close similarity of
curves '1' and '2' can be understood if one notices that the quite
different gap symmetry on the $\gamma$ sheet can lead to rather
similar results after integration over the Fermi surface (Eq. 10).

As a further test of the presence of horizontal line nodes on all
Fermi surface sheets, we have also calculated the in-plane
superfluid density $n_s(T)$ using
\begin{equation}
\frac{1}{\lambda^2(T)}= \mu_0 e^2 \sum_{\rho} \int_{BZ} {\rm d}^3k ~v_i^2 
\left(
\frac{\partial f}{\partial \epsilon_{ {\bf \rm k} \rho}} - \frac{\partial 
f}{\partial 
E_{{\bf \rm k} \rho}} 
\right), 
\end{equation}
where $\lambda(T)$ denotes the temperature dependent penetration depth,
$v_i = v_x$ or $v_y$ is the in plane band velocity at ${\bf \rm k}$,
$e$ is the electron charge, $\mu_0=4\pi \times 10^{-7}$ is the magnetic 
constant, $\epsilon_{ {\bf \rm k} \rho}$ is the electron band energy 
and
\begin{equation}
\frac{n_s(T)}{n_s(0)}= \frac{\lambda^2 (0)}{\lambda^2 (T)}.
\end{equation}

Both
secenarios '1' and '2' give similar results of $n_s$ and agree
with  experimental results by Bonalde {\it et 
al.}\cite{DalevH} (gray squares).
However, one can note that the slightly different slope in small 
temperature regions gives a slight advantage to solution '2' which mimics 
the 
experimental data a little better.

\section{4. Conclusions.} ~\\

We have tested two different gap models for strontium ruthenate,
which are consistent with two physically different pairing
mechanisms. In scenario '1', we assumed that all in-plane
interactions are repulsive, and that only the out of plane
interactions lead to pairing. This is motivated by the Coulomb
scattering pairing mechanism of Koikegami, Yoshida and
Yanagisawa\cite{koik03}.  In scenario '2' we assumed attractive
in-plane interactions for Ru $d-xy$ orbitals ($\gamma$ band) and
attractive inter-plane interactions of the $a$ and $b$ orbitals
($\alpha$ and $\beta$   bands).   Surprisingly we found that the
predicted specific heat is very similar in both models (Fig. 5),
even though one has a horizontal line node on all three Fermi
surface sheets while the other has a nodeless $\gamma$ sheet. 
Similarly the temperature dependent superfuid density is closer to 
experiment in both scenarios.
Of
these two models the nodeless $\gamma$ sheet appears to be
slightly close to the experiment, but the actual differences are
small.

In these calculations we have not attempted to include the
interband proximity effect proposed by Zhitomirsky and
Rice\cite{zhit}. In \cite{annett03} we showed that this
corresponds in real-space to the addition of three-site Hubbard
interaction parameters, or assisted hopping, to the Hubbard
Hamiltonian.  It would be quite possible to combine this interband
proximity effect interaction with either of the two pairing
scenarios which we have considered here. However the close
similarity of the specific heat and superfluid density in the two models, 
shown in Figs.
5, 6 strongly suggests that the effect of the proximity coupling
terms would also be very similar in either pairing model.

\section*{Acknowledgements} ~\\

This work has been partially supported by KBN grant No.
2P03B06225, the NATO Collaborative Linkage Grant 979446 and INTAS
grant number No. 01-654. We are grateful to Prof. Maeno 
and Prof. D. Van Harlingen
for
providing us with the experimental data reproduced
in Figs. 5 and 6.

\end{document}